\documentclass[slac_one]{revtex4}
\usepackage{graphicx}
\usepackage{fancyhdr}
\begin{document}

\title{{\small{Hadron Collider Physics Symposium (HCP2008),
Galena, Illinois, USA}}\\ 
\vspace{12pt}
The ATLAS and CMS Plans for the LHC Luminosity Upgrade} 

%

\author{D. Bortoletto}
\affiliation{Purdue University, West Lafayette, IN 47906, USA}

\begin{abstract}
In January 2007 the CERN director general announced the plan for
the staged upgrade of the LHC luminosity. The plan foresees a phase 1
upgrade reaching a peak luminosity of $3 \times 10^{34}$
cm$^{-2}$s$^{-1}$ followed by phase reaching up to $ 10^{35}$
cm$^{-2}$s$^{-1}$.  We discuss the physics potential and the
experimental challenges of an upgraded LHC running. The detector R\&D
needed to operate ATLAS and CMS in a very high radiation environment
and the expected detector performance are also discussed.
\end{abstract}

\maketitle

\thispagestyle{fancy}


\section{INTRODUCTION} 

The Large Hadron Collider (LHC) at CERN will collide two proton
beams with a center-of- mass energy of 14 TeV (7 times the energy of
the Tevatron's proton-antiproton collisions) at a design luminosity
of $10^{34}$ cm$^{-2}$s$^{-1}$ ($\approx$ 100 times that of the
Tevatron). The machine is expected to lead to the discovery of the
Higgs boson, the last missing block of the standard model of
particle and interactions (SM) and of new particles predicted by
theories going beyond the standard model that should be produced
at these energies.

The first particle beams are expected to be injected in September 2008.
Currently all the LHC magnets have been installed and every
sector has reached the operating temperature of 1.8 K. The initial
collision energy will be around 10 TeV to facilitate the startup. The
full commissioning to 14 TeV is expected to take place after the winter
shutdown.  During the 10 TeV commissioning there will be few colliding
counter-rotating pairs of proton bunches. The bunch crossing time will be
75 ns and the number of proton bunches
will be gradually increased from 43 to 156. The luminosity is
expected to be between $10^{29}$ cm$^{-2}$s$^{-1}$ and $2 \times
10^{31}$ cm$^{-2}$s$^{-1}$. The luminosity ramp up is expected to take
several years. It will start by establishing operation with a 25 ns
bunch crossing time at 14 TeV. Then the number of bunches will increase
from 946 to 2808 yielding a luminosity between up to $1.2 \times$
10$^{32}$ cm$^{-2}$s$^{-1}$. The luminosity will continue to increase
and the machine is expected to reach the design luminosity of $1
\times 10^{34}$ cm$^{-2}$s$^{-1}$ in 2012.

The physics reach of the LHC will be ultimately defined by the
integrated luminosity delivered to the experiments. Exploiting the
full potential of physics at the LHC, which includes upgrading its
luminosity, is the highest priority of the European Strategy for
Particle Physics, which was adopted unanimously by the CERN Council
in July 2006~\cite{eu}. The high scientific priority of this goal
was also highly supported by the US Particle Physics Prioritizing
Panel (P5) in their may 2008 report~\cite{p5}.

The planned upgrade of the machine aims first, in the so called Phase
1, at providing reliable operation at $2 \times 10^{34}$
cm$^{-2}$s$^{-1}$. A decision to implement further machine upgrades to
achieve $1 \times 10^{35}$ cm$^{-2}$s$^{-1}$ will be taken around 2011
once the physics from the LHC data becomes available to guide this
choice. This second phase of the upgrade, often called phase 2 or
SLHC, is especially challenging both for the machine and the
experiments.

Here we discuss the machine plans to increase the LHC luminosity,
the physics reach of an upgraded LHC, and the upgrade needed for the
ATLAS and CMS experiments not only to cope with the more radiation
hard environment but also to maintain or possibly improve  their
physics capabilities. I will describe the current plans developed by
ATLAS and CMS both for Phase 1 and Phase II of the LHC upgrade.

\section{MACHINE UPGRADE}

The LHC luminosity upgrade plan has been under discussion since
2001. Studies on how to improve the integrated luminosity have found
bottlenecks and weaknesses in the proton-injection and accelerator
chain. Significant luminosity gains can be obtained by upgrading the
injector chain which includes elements built in the 50s. Currently
protons are accelerated successively by passing through the Linac 2,
the Booster, the PS and the SPS before being injected into the LHC.

The plans for the luminosity upgrade foresee first a phase 1 that
will reach a peak luminosity of $3 \times 10^{34}$ cm$^{-2}$s$^{-1}$.
The accelerator changes in the LHC injector proposed for 1 phase
include the replacement of Linac 2 with the Linac 4. Currently
proton bunches at 50 MeV are injected from Linac 2 into the Booster
in a process that dilutes the beam brightness. The Linac 4 will
allow acceleration of 160 MeV H$^{-}$ particles followed by
injection in the Booster using a charge-exchange technique that
removes excess electrons. This method will improve beam brightness
and increase the luminosity in the LHC. Plans for the new Linac 4
are well advanced and CERN expects the construction of the machine to begin
soon. The machine will be ready for commissioning by 2012. This will result
in a doubling of peak LHC luminosity. The Phase 1 luminosity
upgrade scenario will also include a new interaction region to reduce
$\beta^{*}$ from 0.5 to 0.25 m. This could also increase the peak
luminosity by a factor 2.  New focusing triplets based on Nb-Ti
superconducting technology are expected to be implemented before the
2013 physics run to achieve this goal. The LHC collimation system and the
separation elements near the interaction regions will also be improved
to complete phase 1 of the luminosity upgrade.

Phase 2 of the upgrade will require further changes in the LHC injector
chain. The Booster and the PS suffer from intensity limitations and have severe
reliability problems because of their aging. Current planning foresee extending
the energy of Linac 4 to several giga-electron-volts with a new machine called
the Low Power Superconducting Proton Linac (LPSPL). This will be followed by a new
50 GeV synchrotron called PS2 which will increase the SPS injection
energy and double the proton flux. The PS2 and the LPSPL are entering the R\&D
and design-optimization stage.  A decision for their construction
will be taken by 2011. The building of these new injectors could place in
parallel with operation of the LHC and they could provide a luminosity
between 8 and $10 \times 10^{34}$ cm$^{-2}$s$^{-1}$ after a one year shutdown
in 2017.

The current expectations for the peak and integrated
luminosity~\cite{garoby} that will be reached by an upgraded LHC are
shown in Table~\ref{luminosity}.


In phase 1 of the upgrade the machine will continue operations at 25
ns bunch crossing which is the LHC baseline. Several different
scenarios for the machine upgrade parameters have been considered to
achieve $1 \times 10^{35}$ cm$^{-2}$s$^{-1}$ in Phase 2. The
original baseline foresaw operating at the ultimate bunch intensity
of $1.7\times 10^{11}$, with 12.5 ns bunch spacing, a crossing angle
about 50\% larger than nominal, and correspondingly bunches of half
the nominal length. The bunches were supposed to be shortened
through a combination of higher harmonic rf and reduced longitudinal
emittance. This scenario was attractive since it could achieve a
luminosity of $1 \times 10^{35}$ cm$^{-2}$s$^{-1}$ but it was
abandoned since the expected heat reaches the maximum cooling
capacity of 2.4 W/m per beam.

At the end of 2006 the Physics Opportunities and Future Proton
Accelerators (POFPA) Committee recommended a 50 ns and 25 ns as the
SLHC baseline and as the backup bunch crossing \cite{pofpa}. The backup
scenario with 25 ns bunch spacing is denoted as the
early-separation(ES) scenario. The beam is squeezed down to a a
$\beta^{*}$ of 10 cm in CMS and ATLAS. The ES scenario adds
early-separation dipoles inside the detectors starting at $\approx
3$ m from the IP and uses crab cavities. This scenario implies
installation of new hardware inside the ATLAS and CMS detectors, as
well as the first ever hadron beam crab cavities. The latter gains a
factor 2 to 5 in luminosity. While this parameter configuration is
less demanding for the injector system modification and it entails
in average 300 pile-up event, it also requires the placement of
dipole magnets adjacent to the inner detectors, which impose severe
constraints for the detector upgrade. In the
50 ns scenario, which is often denoted as the large Piwinski angle
(LPA) scenario longer, longitudinally flat, and more intense bunches
are collided with a large Piwinski angle. The beta function at the
interaction is reduced by a more moderate factor to a $\beta^{*}$ of
25 cm. This scenario does not require accelerator elements inside
the detector. Challenges are the operation with large Piwinski
angle, which is still unproven for hadron beams, the high bunch
charge, and the larger beam current. The LPA scenario is expected to
yield a peak instantaneous luminosity of 1400 pp interactions per
beam crossing (pile-up). The baseline scenario might
require the installation of 8 m long slim quadrupole magnets near
the interaction region, probably between the forward muon wheels.

The total integrated luminosity that will be achieved by the end of phase
I will be about 648 fb$^{-1}$ by the end of 2016. By 2025 the LHC will
deliver about 5028 (2088) fb$^{-1}$ with (without) the Phase 2 upgrade.
The phase 1 luminosity upgrade will already require changes in the detector
elements that will deteriorate because of radiation damage and/or
will be affected by the higher occupancies that complicate pattern
recognition or lead to increased dead time.  Triggers will also
degrade as the increased occupancy reduces the effectiveness of
isolation and correlation algorithms.  The large pile-up and
radiation effect of phase 2 present a major experimental challenge
especially for the detectors elements close to the IP at and in the
forward region. The current ATLAS and CMS plans are discussed.

\begin{table}[t]
\begin{center}
\caption{Peak and integrated luminosity expected between 2009 and
2025 with and without the phase 2 upgrade. }
\begin{tabular}{|l|c|c|c|c|}

\hline &\multicolumn{2}{c|}{\textbf{With Phase 2}}
&\multicolumn{2}{c|}{\textbf{Without Phase 2}}\\ \hline

\textbf{Year} & \textbf{Peak L}  & \textbf{Integrated L}&
\textbf{Peak L} & \textbf{Integrated L} \\

& 10$^{34}$cm$^{-2}$s$^{-1}$& fb$^{-1}$ &10$^{34}$cm$^{-2}$s$^{-1}$ &fb$^{-1}$\\
\hline
2009&   0.1 &   6  &    0.1 &   6\\ \hline
2010&   0.2 &   18 &    0.2 &   18\\ \hline
2011&   0.5 &   48 &    0.5 &   48\\ \hline
2012&   1   &   108&    1   &   108\\ \hline
2013&   1.5 &   198&    1.5 &   198\\ \hline
2014&   2   &   318&    2   &   318\\ \hline
2015&   2.5 &   468&    2.5 &   468\\ \hline
2016&   3   &   648&    3   &   648\\ \hline
2017&   3   &   648&    3   &   648\\ \hline
2018&   5   &   948&    3   &   828\\ \hline
2019&   8   &   1428 &  3   &   1008\\ \hline
2020&   10  &   2028 &  3   &   1188\\ \hline
2021&   10  &   2628 &  3   &   1368\\ \hline
2022&   10  &   3228 &  3   &   1548\\ \hline
2023&   10  &   3828 &  3   &   1728\\ \hline
2024&   10  &   4428 &  3   &   1908\\ \hline
2025&   10  &   5028 &  3   &   2088\\ \hline
\end{tabular}
\label{luminosity}
\end{center}
\end{table}

\section{PHYSICS POTENTIAL}

Current experimental observations predict exciting
discoveries at the "terascale" which will be explored by the LHC.
The physics opportunities at the LHC have been studied over a period
of many years. We expect to finally understand the mechanism by
which electroweak symmetry is broken (EWSB) and the Z and W gauge
bosons acquire their masses. In the Standard Model, this is achieved
by the Higgs mechanism and its manifestation is a scalar particle,
the Higgs Boson, whose mass is expected to be between 0.115
TeV/c$^{2}$ (LEP limit) and 1 TeV/c$^{2}$. The search for the Higgs
Boson is one of highest priority investigations at the LHC and we
expect a discovery within a few years of LHC operations.

The discovery of the SM Higgs will be only begin the Terascale
exploration. Quadratic divergences in the Higgs mass from radiative
corrections lead to the expectation that new physics should appear
near the Terascale. One popular set of BSM physics is
Supersymmetry (SUSY) which solve this problem by introducing
for every SM boson a new SUSY fermion partner and viceversa. These new
symmetry introduces an elegant cancelation of the divergency. If
SUSY is realized in nature we expect exciting discovery of SUSY
particles possibly even before the observation of the several Higgs
particles that are associated with it. The LHC is sensitive also to other
Physics Beyond the Standard Model (BSM) such as Large Extra
Dimensions, new strong dynamics such as Technicolor, new gauge
bosons, or lepto-quarks.

It is evident that more statistics will be required to study in depth
both Higgs physics, rare standard model phenomena, and the BSM. For example if
some hints of SUSY are found, there is a whole set of particles to discover. Only
some of these new particles would be accessible at the LHC. Similarly if one
or more Higgs candidates are discovered, it will be crucial to confirm their
quantum numbers and couplings.  These measurements are necessary to establish if
they are truly elementary objects or are composite particles as expected in
Technicolor.

The luminosity upgrade will provide the statistic necessary for
these measurements and will expand the LHC physics potential. Here
we described the SLHC impact on Higgs and SUSY. More complete
information can be found at~ \cite{manganogianotti}. Nonetheless the
higher luminosity will allow one to push the sensitivity of many
searches to new mass scales especially  $Z^{\prime}$s. One or more
new neutral gauge bosons are highly motivated in beyond the Standard
model scenarios including Higgless theories and large extra
dimensions models. A factor of 10 increase in luminosity extends the
$Z^{\prime}$s reach by 1-1.5 TeV/c$^{2}$.

\subsection{Higgs Physics}

The discovery of a SM Higgs over the full allowed mass range
or of at least one SUSY Higgs boson, will be accomplished at the LHC.  Nonetheless
the luminosity upgrade will enhance the LHC potential by enabling the discovery
of rare
decay modes which extend the information knowledge on the Higgs couplings to
fermions and bosons.

The decay $H \rightarrow Z \gamma$ of a SM Higgs in the mass region
100-160 GeV and has a branching ratio of the order of $10^{-3}$.
Moreover only decays of the Z into electron or muon pairs lead to
final states which can be observed above the background at the LHC.
The production cross-section times branching ratio for $H\rightarrow
Z\gamma \rightarrow \ell\ell\gamma $ is only $\sim$ 2.5 fb yielding
an expected significance for 600 fb$^{-1}$ (300 fb$^{-1}$ per
experiment) of $\ll 3.5$. The factor of 2 increase in luminosity
expected by the end of phase 1 will provide the first measurement
while the factor of ten in luminosity would allow the observation of
a signal at the 11$\sigma$ level. The mode $H \rightarrow \mu\mu$ is
also expected to become accessible at the SLHC and it could be
measured with a precision of 20\%.

\begin{figure*}[t]
\centering
\includegraphics[width=135mm]{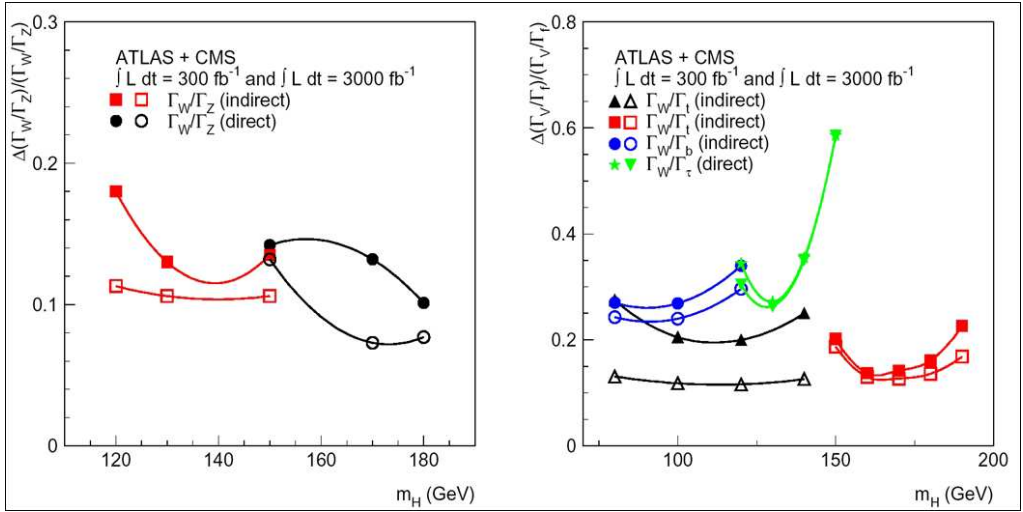}
\caption{Expected sensitivity on the ratios of several Higgs partial
widths for final states involving bosons (left) and bosons and
fermions (right), as a function of the Higgs mass. The closed
symbols correspond to an integrated luminosity of 600 fb$^{-1}$
(LHC: ATLAS+CMS), the open symbols to an integrated luminosity of
6000 fb$^{-1}$ (SLHC: ATLAS+CMS).} \label{higgs}
\end{figure*}

The Higgs couplings to fermions and bosons can determined by
measuring the Higgs production rate in a given channel since
$g_f^{2} \propto \Gamma_f$, where $\Gamma_f$ is the Higgs partial
width for that channel. This requires that the Higgs production
cross-section and total width are known from theory. The ratios of
Higgs couplings, which can be evaluated by measuring the ratios of
rates between two different final states, provides theory
independent measurements since the total Higgs cross-section and
width cancel in the ratio. We expect that at the SLHC ratios of
Higgs couplings to fermions and bosons should be measured with
precisions of 10\% or better in most cases. In some cases, this
represents an improvement by up to a factor of two on the ultimate
precision expected at the standard LHC. Fig. \ref{higgs} show the
SLHC reach. For examples the ratio between the $H\rightarrow ZZ
\rightarrow 4\mu $ and the $H \rightarrow WW \rightarrow \ell \mu
\ell \nu$  which provides a direct measurement of the ratio
$\Gamma_Z/\Gamma_W$, are shown for Higgs masses larger than 150 GeV.
These measurements will provide useful constraints of the underlying
theory.

The SLHC will also open the possibility of measuring the Higgs
self-coupling which probes the non abelian nature of the Higgs
sector. The LHC can not probe this fundamental parameter because of
lack of statistic. At the SLHC we can measure the production of a
pair of Higgs bosons, which is sensitive to the $H \rightarrow HH$
vertex. We expect to reach a sensitivity again of the order of
20-30\%. A promising channel is the $H \rightarrow HH \rightarrow
WWWW \rightarrow \ell\nu\ell\nu jjjj$ decay, with final states
characterized by jets and same-sign leptons.

\subsection{SUSY}

\begin{figure*}[t]
\centering
\includegraphics[width=10cm]{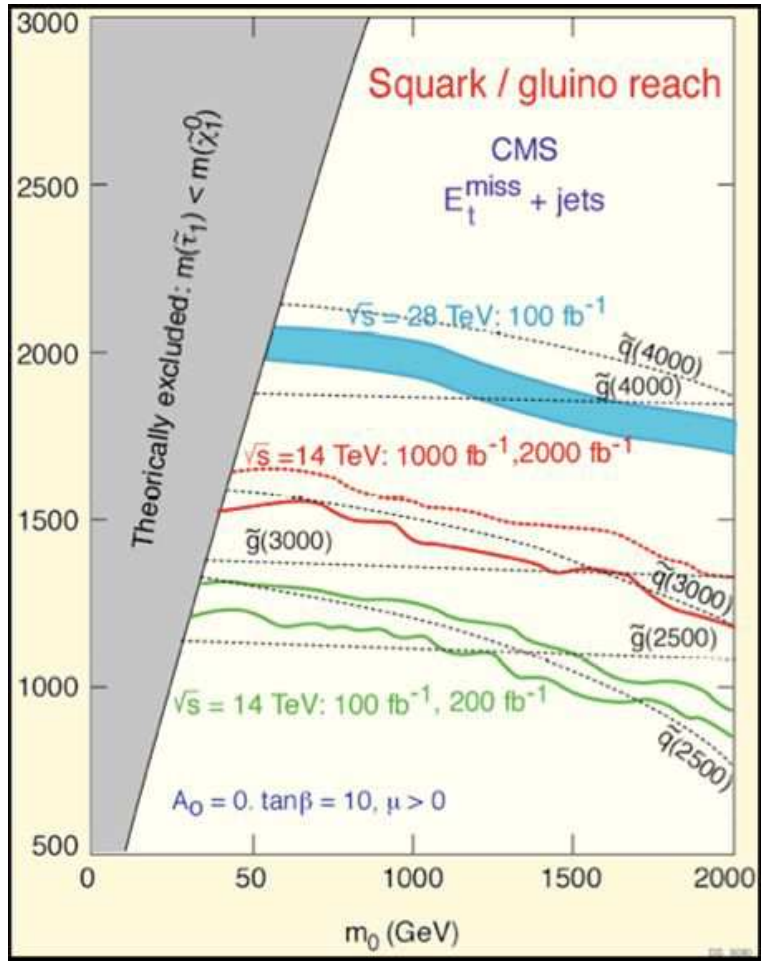}
\caption{CMS 5$\sigma$ discovery reach in the mSUGRA (m$_0$,
m$_{1/2}$) plane for the inclusive multi-jet plus transverse missing
energy final states. The various curves show the discovery reach for
integrated luminosities of 100 fb$^{-1}$ and 200 fb$^{-1}$ (LHC),
for 1000 fb$^{-1}$ and 2000 fb$^{-1}$ (SLHC), and for 100 fb$^{-1}$
and 28 TeV center–of–mass energy (VLHC). Isomass contours for
squarks and gluinos are also shown by the dash-dotted curves.
integrated luminosity of 6000 fb$^{-1}$ (SLHC: ATLAS+CMS).}
\label{susy}
\end{figure*}

ATLAS and CMS are expected to discovery evidence for SUSY if it
exists at the TeV scale. If the LHC delivers a luminosity of 300
fb$^{-1}$ we expect to probe squarks and gluinos masses up to 2.5
TeV/c$^{2}$ as shown in Fig.~\ref{susy}.The various contours are
derived within the framework of minimal Supergravity models
(mSUGRA), and are shown as a function of the universal scalar mass
$m_0$ and of the universal gaugino mass m$_{1/2}$. They were
obtained by selecting events with multi-high-pT jets and large
missing transverse energy, which is most striking SUSY signature if
R-parity is conserved. The SLHC can enhance the discovery reach of
about 0.5 TeV/c$^{2}$ up to 3 TeV/c$^{2}$, as can be seen from
Fig.~\ref{susy}.

Even if SUSY will be discovered at LHC, the reconstruction of SUSY
particles as well as the measurement of model parameters may be
quite difficult at the LHC, depending on the scenario in which SUSY
will manifest itself. The precise measurement of SUSY particles
requires in most cases the selection of exclusive decay modes,
containing e.g. leptons or $b$-jets, and excellent vertexing and
tracking performance must be maintained. Some of these exclusive
channels are expected to be rate-limited at the LHC and would
therefore benefit from a luminosity upgrade. Squark and gluino
reconstruction, becomes very difficult for large $\tan\beta$, the
ratio of the vacuum expectation value of the Higgs
doublets~\cite{tricomi}. In this regime and a large region of the
parameter space will not be available at the LHC.

The LHC luminosity upgrade also increases the reach for the SUSY
Higgs sector. The LHC should be able to discover two or more of the
five SUSY Higgs bosons over most of the parameter space. However in
the region at large $m_A$ which is often defined as the "decoupling
limit", only the lightest state $h$ (which has very similar features
to the SM Higgs boson) can be observed, unless the heavier Higgs
bosons have detectable decay modes into SUSY particles. Neither the
LHC or a future sub-TeV linear collider, could probe heavy SUSY
higgs states. The SLHC, on the contrary, could significantly extend
the region over which at least one heavy Higgs boson can be
discovered covering the full region with $m_A$ $<$ 500 GeV/c$^2$.

\section{ATLAS AND CMS LHC DETECTORS}

The ATLAS detector~\cite{atlas} employs silicon pixels and strips
and a straw-tube based transition radiation detector inside a
superconducting solenoid with a 2 T field. Outside of this, a Pb-LAr
electromagnetic calorimeter is enclosed by iron scintillator
(barrel) and Cu/W-LAr (forward) hadronic calorimetry. The ATLAS muon
system is composed of muon drift tubes, thin gap chambers and
resistive plate chambers embedded in a large array of 8 air-core
toroid magnets.

The CMS detector~\cite{cms}uses silicon pixels and microstrips,
PbWO4 crystal electromagnetic and brass-plastic scintillator
hadronic calorimetry all inside a large 4T superconducting solenoid.
The muon system is composed of drift tubes, cathode strip chambers
and resistive plate chambers inserted between iron layers of the
flux return. Both ATLAS and CMS Level-1 (L1) trigger systems reduce
the input crossing rate of 40 MHz to less than 100 kHz with custom
processing of calorimeter and muon detector information. Further
processing that eventually involves all detector data is used by
their Data Acquisition (DAQ) systems to select an output rate of
$\approx 100$ Hz of data archived.

\section{DETECTOR UPGRADES}

\subsection{CMS}

The CMS plan for the phase 2 upgrade is presented in\cite{eoi}. CMS foresees
to replace several detector elements to maintain the physics
capability for phase 1 of the upgrade. The current
pixel detector with three barrel layers (BPix) and three end-cap disks (FPix)
will be replaced. The possibility to introduce a fourth pixel layer which could be
located at about 16 cm from the interaction region is also under evaluation. The insertion of
this layer could significantly improve tracking if there was
indication that the first silicon strip layer of the Tracker Inner
Barrel (TIB) was degrading more rapidly than expected.

The performance of the current pixel sensors~\cite{sensors}
surpasses the design goals of 6$\times 10^{14}$
neutron-equivalent-particles$/$cm$^{2}$ set in the CMS technical
Design Report~\cite{cms}. Nonetheless it significantly degrades
after $1.2 \times $ 10$^{15}$ neutron-equivalent-particles/cm$^2$
which would be accumulated by about 2014 with the expected ramp up
of the machine. The Phase 1 scenario in table~\ref{luminosity}
indicate that a total dose of $3.2 \times 10^{15}$ $/$cm$^{2}$ 1MeV
neutron equivalent would be accumulated at the inner pixel barrel
layer by 2017 without any upgrade. This will lead to a charge
collection as low as 20 \% towards the end of Phase 1 and to a large
fraction of lost tracks, as shown. The second pixel layer will also
be significantly impacted by the total fluence received by the end
of phase I. The replacement of the pixel detector will allow the
adoption of more radiation hard sensors for the inner pixel layers.
The most promising option is to use n-on-p pixels. Recent results
demonstrate that this technology can operate up to $3\times 10^{15}$
/cm$^{2}$ 1MeV neutron equivalent~\cite{nonp}.

The current pixel readout chip (ROC), known as the
PSI46\cite{psichip}, was originally designed for placement at a
radius of about 7 cm from the interaction region at design
luminosity.  The chip was subsequently improved upon, but
nevertheless there will be a loss of data at a rate of about 4\%
from the readout chip at 10$^{34}$ cm$^{-2}$s$^{-1}$ in the
innermost layer at 4 cm as measured in test beam runs. At the
nominal L1T accept rate of 100 KHz, the data loss of 4\% at
10$^{34}$ cm$^{-2}$s$^{-1}$ will increase to 8\% and could reach
16\% in Phase 1. The data loss is dominated by the buffer size. R\&D
has already started for a new version of the PSI46v2 readout chip
with improved buffering. For a luminosity less than $2.5 \times
10^{34}$ cm$^{-2}$s$^{-1}$ the buffer size should be double to
recover efficiency. Doubling the buffers in the 250 nm process will
increase the size of the chip periphery by 0.8 mm which is
acceptable. The 130 nm deep submicron processes is also under
evaluation.

The luminosity increase expected in Phase II will require the
replacement of the full CMS tracking system. The new tracker will be
installed during a one year shutdown before the start of Phase 2.
CMS is developing a strawman layout which provides excellent
tracking and vertexing while also allowing tracking information for
L1 triggers. The present CMS tracker produces a large amount of
information for each Level 1 Trigger, at up to 100 KHz. For the
Strip Tracker, the analogue information for each of the
approximately 9 Million strips is read out for every triggered
event. For the pixel detector, zero suppression is used to read out
only the information for clusters above threshold. In the high hevel
trigger (HLT), only clusters within given regions of interest are
considered, at least in the early stages of the selection. It is
implausible to access all this information for the purposes of a
Level 1 Trigger at the SLHC and zero suppression will be required
throughout the upgraded Tracker. Additional data reduction of at
least one to two orders of magnitude is required prior to
transmission for the L1 trigger decision. CMS is considering
measuring track vectors with pairs of silicon sensors separated by a
few centimeters to identify tracks with $p_T$ above 10-20 GeV.
Detailed studies are underway, to validate the performance of this
layout and optimize its design~\cite{trig1}.

Another approach to reduce the Level 1 trigger relies on associative
memories chips and radiation hard optical fiber technologies. One of
the proposed systems is an evolution of the Silicon Vertex Tracker
(SVT) built for the CDF experiment, at the Tevatron Collider, where
tracks are reconstructed in real time. In order to cope with the
large data volume to be processed, CMS plans to use the information
coming from the radial region from 25 to 50 cm where radiation
levels allow the operation of current hybrid pixel technologies and
optical fibers. The large radius would also provide the lever arm
required for a momentum measurement in the $r-\phi$ plane. Studies
are taking place to evaluate if such an approach can achieve the
needed data reduction~cite{trig2}.

The upgrade of the hadronic calorimeter (HCAL) foresee replacing the
current hybrid photo-detectors (HPDs) with Silicon Photo-multiplier
(SiPM) which have lower noise and higher gain.  SiPMs been shown to
work at magnetic fields between 0T and 4T with higher quantum
efficiency than HPDs and have a very large gain of ~10$^{6}$.  The
small size and low cost of the SiPm  will allow depth segmentation
which is not possible with the current hybrid photo-detectors (HPD)
due to space constraints. Introducing longitudinal segmentation will
allow to correct for the loss of light in the inner layers of the
Hadron Endcap calorimeter (HE) due to radiation damage. The large
gain will also permit accurate timing measurements of the deposited
energies. This could be especially important since the increased
out-of-time pileup at higher luminosity will degrade the jet and
missing transverse energy (MET) resolution. These effects will
introduce significant systematic uncertainties for the jet energy
scale, and threaten the use of lepton isolation in both the hardware
and software triggers.

New trigger boards will needed to accommodate the changes in the
front end electronics due to the increased fiber data bandwidth and
number of channels. We are closely evaluating the performance of the
$\mu$TCA which incorporates the latest trends in high speed
interconnect and switching, and takes advantage of industry
standards.  In $\mu$TCA, crates the controllers are accessed over
ethernet, and the backplane consists of a large number of high speed
(1-5 Gbps) serial links. High-speed cross-point switches could allow
point-to-point communications and therefore yield the flexibility
needed to combine information from various detectors. These
correlations, which are not permitted in the current trigger
architecture, could be essential for implementing more sophisticated
trigger algorithms to cope with the higher luminosities.

CMS is also planning changes to the ECAL trigger system. Closer
coupling of the ECAL and HCAL off detector electronics will yield a
more consistent treatment of calorimeter energies associated with
physics objects in the trigger/DAQ. ECAL trigger primitives could be
sent to HCAL to assure all corresponding HCAL towers are read out
for analysis of EM objects. Specifically, once the longitudinal
segmentation in HCAL is increased, it will be advantageous to ensure
that the first layers of HCAL are always read out to contain leakage
for high energy EM objects and for use in isolation. Similarly,
trigger primitives or regions of interest in HCAL should initiate a
full readout of corresponding ECAL towers to ensure that cells
associated with jets and taus are available at high level trigger
and off-line. Such functionality will be relevant to maintaining
consistent quality of physics objects with increases in default
readout thresholds.

Further upgrades to the calorimeters might be needed for Phase 2.
The increase in pile-up events and the harsher environment in the
forward region requires a carefully study of the radiation
resistance of the CMS endcap crystals. The scintillators of the CMS
hadronic endcap calorimeter will likely be replaced.

\begin{figure*}[t]
\centering
\includegraphics[width=6cm]{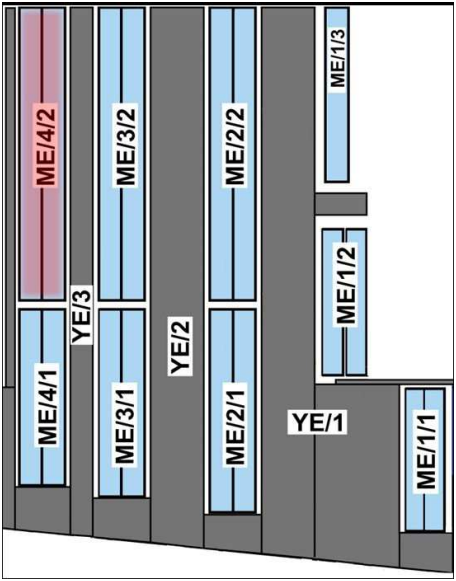}
\hfill
\includegraphics[width=8cm]{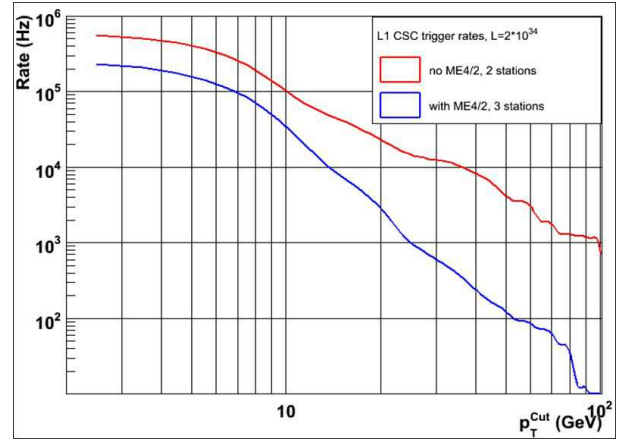}
\caption{ Left: Cross-section of one quarter of the CSC muon system
showing the placement of proposed ME4/2 chambers, and other CSC muon
stations including ME1/1. Right:SC muon trigger rate curves without
ME4/2  (upper) and with ME4/2 (lower). The thresholds to achieve an
acceptable rate at $2\times 10^{34}$ cm$^{-2}$s$^{-1}$ of 20 KHz are
$\approx$ 20 GeV and 60 GeV with and without ME4/2 respectively.}
\label{muons}

\end{figure*}

The CMS muon system includes drift tube (DT) in the barrel region,
cathode strip chamber (CSC) in the endcap and resistive plate
chambers (RPC) in both the barrel and endcap. The barrel region (0
$<\eta<$ 1.2) is composed by 5 wheels, each divided in 12 sectors
with 4 iron gaps. The muon stations consist of one DT chamber and
two RPC chambers joint together and are placed in the gaps of the
iron return yoke plates. The two endcap are made of 3 iron disks and
4 layers divided in 2 or 3 stations (ME1/1, ME1/2, ME1/3, ME2/1,
ME2/2, ME3/1, ME3/2, ME4/1 and ME4/2) of CSCs and RPCs (see Figure
~\ref {muons}). Excellent trigger performances on single and
multi-muons events and an unambiguous identification of the bunch
crossing is obtained by combining the RPC which are fast and
dedicated trigger detectors with detectors having precise spatial
resolution like the DT and CSC. DT and CSC first process the local
information of every chamber generating local triggers then muons
from different chambers are collected by the Track Finder which
combines them to form a muon track with an assigned transverse
momentum value. The highest $p_T$ muon candidates from each system
are selected and sent to the Global Muon Trigger which applies
transverse momentum thresholds.

The CSC and the RPC are expected to survive the increased radiation
levels from the phase 1 LHC luminosity upgrades. However, the muon
triggering and readout require specific upgrades in order to
maintain excellent triggering characteristics, high efficiency, and
excellent position and time resolution despite the increased
background levels. CMS proposes to recover triggering capability by
adding the ME4/2 chambers. This addition will be necessary for Phase
1 to obtain an efficient L1 trigger threshold in the rapidity range
1.2 to 1.8 as shown in Fig.~\ref{muons}. At high luminosity, the
rate of low-momentum muons becomes too high to run the CSC Track
Finder in a two out of three station triggering configuration, and
another CSC chamber station is needed to allow a three out of four
station triggering configuration. CMS is also planning to upgrade
the ME1/1 which is critical for muon momentum resolution, but
suffers high particle rates which are at the limit of the data
acquisition system to handle.  For Phase 1, CMS will replace the
current cathode front-end boards with digital boards that
flash-digitize every channel rather than using a custom analog
storage pipeline. The smaller size of the new boards should allow
placing seven of these boards on each ME1/1 chamber, and thereby
recover muon trigger capability at higher luminosity in the rapidity
range 2.1 to 2.4. In order to read out the new board will require
some revision of the trigger electronics.

CMS is also planning to replace the muon port cards with high speed
optical links and FPGAs, since they are currently the major CSC
trigger data bottleneck, allowing only three muon stubs to be
transmitted per crossing per 60-degree trigger sector. This will
cause severe problems at the Phase 1 luminosity.  CSC chambers are
also equipped with trigger electronics to find muon track segments.
A Local Charged Track (LCT) is formed when at least 4 out of the 6
layers give signals, either the strips (CLCT) or the wires (ALCT),
which line up in a pattern consistent with a particle going through.
The ALCT cards are currently the largest source of deadtime to the
entire CMS experiment caused by periodic resets that recover from
neutron-induced single-event upsets. These boards will also be
replaced before Phase 1 of the luminosity upgrade.

The Muon systems should not be greatly affected by the Phase 2
luminosity increase. However, due to the modification to the LHC
beam line, the forward shielding will probably need to be increased,
thereby reducing the spectrometer acceptance from about $|\eta| <$
2.5 to about $|\eta|  < $2.0.

The CMS trigger system will need modifications to operate with
adequate performance at the LHC Phase 1 upgrade luminosity of 2-4
$10^{34}$ cm$^{-2}$s$^{-1}$.  For Phase 1, the new information in
the L1 trigger will come from use of more fine-grained information
from the calorimeter and forward muon triggers and improved
algorithms exploiting this new information. Further trigger upgrades
will be needed for Phase 2. The Level-1 Trigger output should not
exceed the 100 KHz limit in order to avoid rebuilding the front end
electronics where possible.  CMS expects that to deal with the SLHC
luminosity of $10^{35}$ cm$^{-2}$s$^{-1}$ tracking information will
be needed at L1.

An important feature of CMS that facilitates the incremental upgrades
foreseen for Phase 1 is the relative accessibility of the parts of
the detector that need to be modified. The End Cap Muon detector
disks can be separated for maintenance and the surface on which the
ME4$/$2 chambers mount is readily accessible. This will allow CMS to
use several shutdowns for their installation.  The Readout Boxes of
the HCAL are similarly accessible once the CMS detector is opened.
The Pixel Detector can be removed in two days (after time has been
allowed for the radiation level to decrease and CMS has been opened)
and a new one installed, cabled, and checked out in a few weeks. The
underground control room that houses much of the Level 1 Trigger and
DAQ can be accessed even while the beams are colliding.

\subsection{ATLAS}

The ATLAS collaboration is also focusing on the detector upgrades for the
SLHC. ATLAS is planning modifications to their beampipe. Currently
central part of the ATLAS beampipe is made of beryllium (Be), while the
remaining part is stainless steel which can get activated and can yield
large backgrounds, especially to the muon system.  The beam pipe will be
changed to Al for phase 1 and the Be for SLHC. This will lead
to a large background reduction in critical areas of muon system.

The most significant upgrade for ATLAS will be the full replacement of
the tracker systems, which will be showing signs of serious
radiation damage by 2016.  Many changes are needed, including the
replacement of the whole Inner Detector since the current TRT will
not cope with the high hit rate. The Pixels and
parts of the semiconductor tracker (SCT) will also have to be substituted since
they will have suffered significant radiation damage.
Moreover the SCT strip occupancy in
certain regions would be very high at the peak luminosity of the LHC, degrading
tracking performance.

The target survival for operation in the
intense radiation field of the SLHC is assumed to be 6000 fb$^{-1}$. To
cope with the severe increase in integrated radiation dose and the
much higher track density (dN$_{ch}$/ d$\eta \approx$ 1500 from up
to 200 events/crossing), the data links, data acquisition,
triggering, off-detector electronics and computing will all need to
be reworked across much of the experiment.

\begin{figure*}[t]
\centering
\includegraphics[width=13cm]{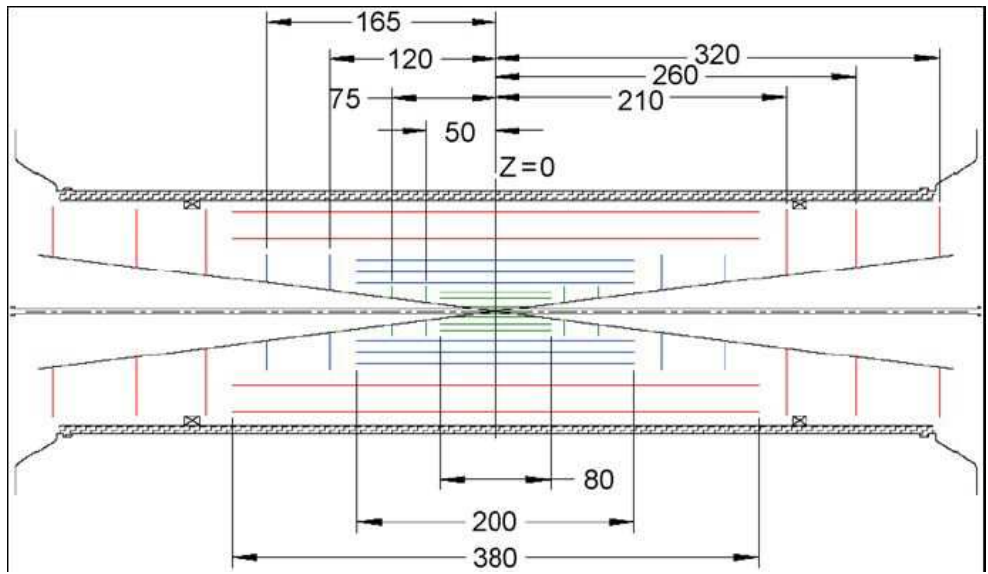}
\caption{View of a basic all-silicon layout for a super-LHC tracker.
Outer two layers contain long (10$-$12 cm) strips. Intermediate
three layers contain short (3 cm) strips. Innermost layers are
pixels.} \label{layout}
\end{figure*}

The Atlas collaboration has developed a strawman tracker design
which is a first step towards establishing a baseline design. The
process is driven by the need to achieve the necessary radiation
hardness and the granularity to handle the increased occupancy
rates. The ATLAS inner detector (ID) will need to be replaced since
the upgraded luminosity will lead to doses greatly exceeding its
design radiation hardness (730 fb$^{-1}$ integrated luminosity).
More detector channels are needed to maintain good pattern
recognition performance, but without exceeding the limited power and
material budgets. Simulations are underway to investigate the
performance of different tracker layouts~\cite{allp}. The expected
requirement of 1\% maximum occupancy for adequate pattern
recognition puts constraints on the technology that can be used at a
certain radius. The results showed that adequate pattern recognition
could be achieved with three pixel and four short strip (3 cm long,
80 $\mu$m pitch) layers. The TRT will also be replaced by two layers
of long strip (9 cm length, 80 $\mu$m pitch) sensors giving
approximately the same position resolution as the TRT. This layout,
shown in Figure~\ref{layout} with 9 silicon tracking layers from 5
to 95 cm radius~\cite{allp} guarantees that the performance at SLHC
should be similar to that of the current ATLAS tracker. The increase
in sensor radiation damage and the required increase in electronic
channel density also lead to major challenges in terms of power
supply and cooling.

A major inter national effort is already underway to develop the
technologies required for the tracker replacement. Radiation
exposure concerns dominate the pixel region. For the innermost
$b$-layer the levels of radiation background are so high that a new
technology is required. ATLAS has a vibrant R\&D program examining
the use of diamond, which is a very radiation hard material, to
build the inner layers for the SLHC.  The ATLAS R\&D program has
already achieved major milestones including  building and
successfully operating in a beam test a full-sized ATLAS pixel
diamond detector. A signal-to-noise ratio of 24 at a field of  2V/mm
was obtained with a diamond sensor irradiated up to 1.8 $\times$
10$^{16}$ p/cm$^{2}$, which corresponds to the 5-year integrated
radiation dose at the SLHC~\cite{diamond}. Another potential
technology for the inner layers is the 3D geometry which has shown
excellent radiation hardness. In a 3D silicon detector, arrays of
$n$ and $p$ electrodes are implanted in a silicon substrate. The
advantages of this technology over the standard planar technology
are: shorter collection distances, faster collection times, better
signal efficiency and lower depletion voltages. Tests with 3D
sensors bump-bonded to an ATLAS pixel readout chip have been
performed in the CERN H8 pion beam. Preliminary results indicates
that 3D sensors might lead to a better signal efficiency than
diamond. The production of 3D sensors is currently using in-house
processing but ATLAS is working with vendors to industrialize the
process\cite{3d}.

The sensor requirements at the intermediate radius are the focus of
much of the ATLAS simulation effort.  The bulk of the tracker will
be silicon planar technology. Silicon n-in-p is favored compared
with p-in-n because it can operate partially depleted, avoiding the
need for very high bias voltages. The ATLAS collaboration is also
developing the concept of a stave, an integrated multi-module object
which is the basis for this new design. A stave is a self-contained
object which includes a number of individual modules sharing a
common mechanical, thermal, and electrical service
structure~\cite{stave}. The stave consists of a core fabricated as a
sandwich of high modulus carbon fiber laminates around a spacing
material such as foam or honeycomb. This sandwich construction leads
to high stiffness against gravitational sag.  Within the spacing
material are a U-shaped cooling loop which should offer the
excellent cooling performance needed for operation a high radiation
environment. Both mechanical and thermal simulations and electrical
tests of stave structures show promising results.  A full scale
ATLAS stave design will be fabricated and tested in the near future.
The results of these studies will guide the design an all-silicon
tracker for the SLHC. The replacement of the full tracker system
will take place during the one shut down before phase 2 of the
upgrade. ATLAS is also discussing a plan to replace the $b$-layer,
before the phase 1 upgrade.

Most of the ATLAS calorimetry should be robust against the increased
radiation backgrounds, although R\&D is required in some areas.
ATLAS is evaluating a novel SiGe Bi-CMOS technology for the upgrade
of the LaR readout.  A baseline that allows operating at the rate
and radiation hardness expected at the SLHC requires the
digitization of all signals as soon as possible. Therefore the
collaboration foresees using very fast ADC on the front end board
(FEB) to avoid using an analog pipeline on the detector.  To achieve
this goal, the power lost in the digitization must be considerably
reduced, possibly at the level of 0.4 W/ADC. Then the pipeline could
be 'off detector' if fast optical link could be used.

The ATLAS forward calorimeter may also need upgrading due to possible
beam heating of the LAr. If its functionality is compromised then
drastic solutions such as a new warm forward calorimeter in front of
the existing one may have to be considered. A safety factor of five
was included in the design. If the predictions are accurate it is
estimated that only parts of forward chambers would need to be
replaced with chambers of higher rate capability. However, if the
background predictions are underestimated by a factor of five then
most of the chambers would have to be replaced. Experience with
running at the LHC is necessary before identify the best solutions.

The Monitored Drift Tube (MDT) chambers and Cathod Strip Chambers
(CSC) in the small angle region are the main tools for the muon
identification and measurement in the ATLAS muon spectrometer. The
main background for the muon chambers are low energy photons and
neutrons which dominate the counting rate in most areas of the
spectrometer, where an overall maximum counting rate of 500
Hz/cm$^{2}$ is expected. The upgrade to SLHC will involve fluxes ten
times higher and therefore testes were carried out to understand the
performance of the chambers after intense neutron dose. The
irradiation of the MDTs was carried out at the ''Tapiro'' nuclear
reactor facility, at ENEA Casaccia laboratories, on a MDT test
chamber~\cite{atmu}. The main goals of the studies  was to
understand the MDT performance under ATLAS-like neutron rates, and
to test chamber robustness after an integrated neutron flux
corresponding to $\approx$ 40 years of real data taking. Cosmic ray
data were acquired and analyzed to look for possible loss in gas
gain or tracking efficiency. No significant variation from the
standard MDT behavior was observed, neither at high background
neutron rates, nor after massive irradiation.

ATLAS Level-1 muon trigger is provided by Resistive Plate Chambers
(RPC) in the barrel and Thin Gap Chambers (TGC) in the end-cap.
Ageing studies have been conducted and the preliminary results show
that the RPC operate at SLHC if the background corresponds to
simulation and the expected background can be reduced by a factor 2
by proper shielding. ATLAS expects that the first 1-2 year of
operation are necessary to draw realistic conclusions on the
behavior of the RPCs at the SLHC. Because of these uncertainties
ATLAS is evaluating and upgrade of the muon forward region. An
option under consideration is to upgrade forward trigger chambers
(TGC) and possibly replace some forward MDT-CSC by smaller diameter
tubes. Micromega chambers~\cite{micro} are a possible solution for
precise measurement and trigger at the same time, due to their rate
capability (tens of kHz/cm$^{2}$), time resolution (few ns) and space
resolution (better than 100 $\mu$m). An R\&D effort on Micromega
chambers has been started by the ATLAS collaboration.

\section{CONCLUSIONS}

The LHC is expected to open the window to the Terascale physics.
Nonetheless a deeper understanding of electroweak symmetry breaking
and physics beyond the standard model is likely to require large
amount of data. A luminosity upgrade of the LHC leads to the full
exploitation of the LHC physics potential. The SLHC will extend the
discovery mass reach by 20 to 30 \% and improve the sensitivity for
precision measurements, for a modest capital investment compared to
the LHC cost. Operation at the very high luminosity of the  SLHC
poses significant challenges to the ATLAS and CMS experiments. R\&D
is started both for the phase 1 and phase 2 luminosity upgrades.

\subsection{Acknowledgments}
The author wish to thank the organizers for the excellent conference
and many members of the CMS and ATLAS collaborations for the input
provided in preparing this talk.

This work is supported by Department of Energy contract
DE-FG02-91ER40681A29 and by NSF funding for the maintenance and
operation of CMS.

\end{document}